\documentclass[%
 reprint,
 amsmath,amssymb,
 aps,
]{revtex4-1}

\usepackage{amsmath}
\usepackage{graphicx}
\usepackage{dcolumn}
\usepackage{bm}

\begin{document}

\preprint{APS/123-QED}

\title{Limits on Observation in Quantum Gravity and Black Holes}
\author{George Alexander Davila}
 \email{Geo.A.Davila@GMail.com, GDavila@Fordham.edu}
\affiliation{%
Department of Physics and Engineering Physics \\
Fordham University \\
Bronx, N.Y. 10458
}%

\begin{abstract}
We discuss how the bounds on observation associated with the Planck units would affect an observers’ perception of a black hole. By simply imposing Planck scale quantities as the lower bounds for length, time, and mass of black hole formation, interesting insights into the nature of black holes can be gained. We also see the emergence of a Planck-scale mass that plays an important role in the observation of black holes and the emergence of a new mechanism for virtual black hole formation. 
\begin{description}
\item[PACS numbers]
04.20.-q, 04.60.-m, 04.70.-s, 04.70.Dy
\end{description}
\end{abstract}

\pacs{Valid PACS appear here}
\maketitle


\section{\label{sec:level1}Introduction}

Using the Schwarzschild metric for a stationary observer outside a black hole, we let:
\begin{equation}
\tau = t\sqrt{1-\frac{r_s}{r}}
\end{equation}

Where $\tau$ is the proper time, t the inertial time, $r_s$ the Schwarzschild radius, and r the radial coordinate. We will restrict the proper time to the Planck time $t_p$ , what is thought to be the smallest observable time within the context of Quantum Gravity [1], and restrict the inertial time to the lifetime of the black hole $T_{BH}$, what is in essence the longest one can observe a black hole (it is assumed that the black hole is isolated and therefore cannot gain mass). The distance of an observer from a black hole will be isolated by separating the radial coordinate into the Schwarzschild radius and the distance from the event horizon. This distance will then be compared to the Planck length $L_p$, the smallest length scale in the context of quantum gravity [2][3][4], in order to measure the proper time at this distance. How these methods relate to each other will be shown in Section 3. Only black holes with mass M $\geq$ $m_p$ where $m_p$ is the Planck mass will be considered for the purposes of this paper, as this is thought to be the lower bound on black hole mass [5].

\section{\label{sec:level2}Application of Planck Scale Bounds}
Let $\tau = t_p = \sqrt{G\hbar/c^5}$ be the Planck time and $t=T_{BH}=(G^2 M^3)/(\hbar c^4 )$ the lifetime of a black hole of mass M, then:
\begin{equation}
t_p=T_{BH} \sqrt{1-\frac{r_s}{r}}
\end{equation}
Which, by moving G, c and $\hbar$ to the left side, can be turned into ${m_p}^6=M^6 (1-\frac{r_s}{r})$ Where $m_p=\sqrt{\hbar c/G}$. r can be expressed as:
\begin{equation}
r=r_s (1-\frac{{m_p}^6}{M^6})^{-1}	
\end{equation}
Since r is the radial coordinate outside the black hole and includes the Schwarzschild radius, we can write:
\begin{equation}
r=r_s+x
\end{equation}
Where x is the distance of the observer from the Schwarzschild radius of the black hole, rather than its center, and: \\

$r = r_s (1-\frac{{m_p}^6}{M^6} )^{-1}=r_s+x$ \\

$x=r_s (1-\frac{{m_p}^6}{M^6})^{-1} - r_s= r_s [(1-\frac{{m_p}^6}{M^6} )^{-1}- 1 ]$ \\

Which becomes:
\begin{equation}
x=r_s (M^6/\frac{{m_p}^6}{M^6} )^{-1}
\end{equation}

Setting $\hbar$=c=G=1, 
\begin{equation}
x=2M(M^6-1)^{-1}=\frac{2M}{M^6-1}
\end{equation}
Which is positive for all M $>$ 1 and negative for 1 $>$ M $>$ 0

This quantity, x, represents the distance an observer needs to get from the Schwarzschild radius to experience a proper time within the lifetime of the black hole, i.e. an observer at x will witness the black hole as having lifetime $t_p$. Since $t_p$ is the smallest observable time within the context of Quantum Gravity, x is therefore the closest an observer can get to a black hole of mass M and still observe the black hole. Subtracting the Planck length $L_p=\sqrt{G\hbar/c^3}$ from (6), using $\hbar=c=G=1$,
\begin{equation}
x-L_p=2M/(M^6-1)-1
\end{equation}

$x-L_p \geq 0$, because the Planck length is the smallest length one can observe.
$x-L_p=0$ for $M \approx 1.22981$, or $M \approx 1.22981m_p$ $(m_p=1)$. Let $\beta \approx 1.22981$ represent this numerical constant, such that $x-L_p = 0$ for $M= \beta m_p$. For $M > \beta m_p$, $x-L_p < 0$, and for $m_p \leq M< \beta m_p$, $x-L_p > 0$, meaning that $x=\frac{2M}{M^6-1}$ only acts as a lower limit on distance for $m_p \leq M < \beta m_p$. It will be shown that the Planck length acts as the lower bound on x for any $M > \beta m_p$. Take x without having $\tau = t_p$:
\begin{equation}
x=r_s (\frac{T_{BH}^2}{\tau^2}-1)^{-1}
\end{equation}
So that x is simply the distance from the Schwarzschild radius for a black hole with lifetime $t=T_{BH}$. Now let $x=L_p=r_s (\frac{T_{BH}^2}{\tau^2}-1)^{-1}$
Isolating $\tau$:
\begin{equation}
\tau=t_p  \frac{M^3}{m_p^3}  (\sqrt{\frac{2M}{m_p} +1})^{-1}
\end{equation}
Letting $\hbar$=c=G=1:
\begin{equation}
\tau =M^3 (\sqrt{2M+1})^{-1}=\frac{M^3}{\sqrt{2M+1}}
\end{equation}
Here $\tau$ represents the proper time for an observer at distance $L_p$ from the Schwarzschild radius. Since $L_p$ is the smallest observable length, $\tau$ corresponds to the smallest observable time. So an observer at a distance of one Planck Length for a black hole of $M > \beta m_p$ would see the black hole dissipate in a time $\tau=\frac{M^3}{\sqrt{2M+1}}$. This expression does not affect the validity of (6) since (10) is only valid for $M> \beta m_p$, as will be shown. Subtracting the Planck time $t_p$ from (10), using $\hbar$=c=G=1:
\begin{equation}
\tau-t_p=\frac{M^3}{\sqrt{2M+1}}-1
\end{equation}
$\tau-t_p \geq 0$, because the Planck time is the smallest observable time. $\tau-t_p=0$, for $M=βm_p$ ($m_p=1$), where $\beta = 1.22981$ is the same constant mentioned earlier. Additionally $\tau-t_p>0$, for all $M> \beta m_p$.
	Setting equation (6) equal to (10) $x=\frac{2M}{M^6-1}=\tau=\frac{M^3}{\sqrt{2M+1}}$ the equation for the mass becomes $M^{16} - 2M^{10} + M^4 - 8M + 4=0$, which only has one solution that is real and positive, which is $M \approx 1.22981 = \beta$, as would be expected.

\section{\label{sec:level3}Analysis}
Letting $\hbar=c=G=m_p=L_p=t_p=1$, it can be extrapolated that: \\

For $m_p \leq M < \beta m_p$: \\
		$x=\frac{2M}{M^6-1}$ acts as the lower limit for the distance an observer can get from a black hole of mass M. This corresponds to a proper time $\tau=t_p$. An observer cannot get within a Planck length of a black hole for this mass interval. \\
		
For $M=\beta m_p$: \\
		An observer can get as close as one Planck length $L_p$ away from the Schwarzschild radius and witness the black hole to have a lifetime equal to the Planck time $t_p$, i.e. (6) becomes $x=L_p$, and (10) becomes $\tau=t_p$ and both hold.\\
		
For $M>\beta m_p$: \\
		$\tau=\frac{M^3}{\sqrt{2M+1}}$ acts as the lower limit for the proper time witnessed by an observer at a distance of one Planck length $L_p$ away from a black hole of mass M. An observer can get as close as the Planck length $L_p$ from the horizon along this mass interval. \\

\section{\label{sec:level4}Discussion}
It is clear that although quantizing length and time eviscerates the continuity of equation (6) and (10), there are still clearly defined values for proper time and distance from the Schwarzschild radius for all $M>m_p$. A mass scale of $M= \beta m_p$ also seems to play an important factor as a cutoff value.

For $m_p \leq M< \beta m_p$, many black holes may be entirely unobservable for many observers. Any observer closer than $x=\frac{2M}{M^6-1}$ for a black hole of mass M would be unable to observe the black hole since the observers’ proper time with respect to the black hole would be less than the Planck time. Since the limit of $x=\frac{2M}{M^6-1}$ from the right hand side approaches infinity as $M \rightarrow m_p = 1$, this can include a wide range of masses. For instance, a black hole of mass $M=(1+{{10}^{-35}}) m_p$ would have a $x=2({10}^{175} ) L_p \approx 3.3({10}^{140})$ meters, a distance much larger than the size of the observable universe. Observers at a distance less than $x=\frac{2M}{M^6-1}$, could not observe such black holes, but they could still exist for such an observer in a fleeting manner. This would allow virtual black holes [6][7] to form in intervals less than the Planck time for particular observers, providing another method of virtual black hole formation within the context of general relativity with only basic quantum gravitational constraints. 

Furthermore, different observers may observe the existence of different black holes. Consider two different observers around a black hole of mass M, $m_p \leq M < Bm_p$, Observer A with $x_A \geq x$, Observer B with $x_B<x$; Observer A would see a black hole while Observer B would not. If this is applied to a large number of black holes, one observer can observer a significantly different mass than another observer without the observers being very far from each other. For instance, if Observer A saw N black holes each of mass (or average mass) M, $m_p \leq M<Bm_p$, and Observer B saw none, then A would see NM more black hole mass than B, where B would see this mass as its coinciding hawking radiation. For sufficiently large N, this could cause different observers to see markedly different versions of the same portion of space. Although Observer B could not directly observe these black holes, B could observe them indirectly by observing how observer(s) A with $x_A \geq x$ interacts with the black holes or observing the resulting Hawking radiation. This effect would not interfere with mass-energy conservation, as B would see the mass of these black holes emitted as their resulting Hawking radiation.

In conclusion, by imposing the Planck scale as a lower bound for equations in classical general relativity we see the emergence of a number of properties previously unknown in the context of General Relativity.

\end{document}